%% file: main_Eucap_final.tex
\documentclass[a4paper,10pt]{IEEEtran}
\renewcommand{\baselinestretch}{0.99}
\usepackage{hyperref}
\usepackage{cite}
\usepackage{amsmath,amsthm,amssymb,amsfonts,steinmetz}
\usepackage{graphicx,flushend}

\def\BibTeX{{\rm B\kern-.05em{\sc i\kern-.025em b}\kern-.08em
    T\kern-.1667em\lower.7ex\hbox{E}\kern-.125emX}}

\usepackage{tikz}
\usetikzlibrary{calc}
\makeatletter
\newcommand{\gettikzxy}[3]{%
  \tikz@scan@one@point\pgfutil@firstofone#1\relax
  \edef#2{\the\pgf@x}%
  \edef#3{\the\pgf@y}%
}
\makeatother

\usepackage{comment}
\usepackage{algorithm,algorithmic}
\usepackage{subcaption}

\input{./Definitions.tex}

\DeclareMathAlphabet\mathbfcal{OMS}{cmsy}{b}{n}

\begin{document}
%\title{Dynamic Metasurface-Based Antenna Localization: EM-Compliant Modeling and Optimization}
%\title{Physically Consistent Optimization of\\ Dynamic Metasurface Antennas for Bistatic Sensing}
%\title{2D Waveguide-Fed Metasurface Antennas: Coupled-Dipole Modeling and Bistatic Sensing}
%\title{Two-Dimensional Metasurface Antenna Arrays: Modeling and Optimization for Bistatic Sensing}
\title{2D Waveguide-Fed Metasurface Antenna Arrays: Modeling and Optimization for Bistatic Sensing}
\pagenumbering{gobble}
\author{\IEEEauthorblockN{
Ioannis Gavras$^1$, Panagiotis Gavriilidis$^1$, and George C. Alexandropoulos$^{1,2}$ 
} 
\\
\IEEEauthorblockA{$^1$Department of Informatics and Telecommunications, National and Kapodistrian University of Athens, Greece}
\IEEEauthorblockA{$^2$Department of Electrical and Computer Engineering, University of Illinois Chicago, USA}
\\
\IEEEauthorblockA{emails: \{giannisgav, pangavr, alexandg\}@di.uoa.gr}
\vspace{-1.1cm} 
\thanks{This work has been supported by the SNS JU project TERRAMETA under the EU's Horizon Europe research and innovation programme under Grant Agreement number 101097101, including also top-up funding by UKRI under the UK government's Horizon Europe funding guarantee.}}

\maketitle
\begin{abstract}
This paper presents a physics-consistent framework for bistatic sensing incorporating a $2$-Dimensional (2D) waveguide-fed metasurface antenna array capable of realizing eXtremely-Large Multiple-Input Multiple-Output (XL MIMO) apertures. A coupled-dipole model is presented that captures the array's mutual coupling due to both waveguide and free-space interactions, and a novel passivity constraint on the corresponding magnetic polarizabilities is proposed. Focusing on a bistatic sensing setup, we leverage a Neumann-series approximation of the array response model and derive the Cram\'{e}r-Rao bound for multi-target parameter estimation, which is then incorporated into a sensing optimization formulation with respect to the metasurface's per-element resonance strength configuration. Simulation results on the position error bound in the radiative near field with the proposed design quantify the critical role of metamaterial placement in strongly coupled metasurface-based XL MIMO bistatic sensing systems.
\end{abstract}
\vspace{-0.2cm}

\vspace{-0.3cm}

\section{Introduction}
Metasurfaces constitute a class of reconfigurable antenna panels composed of densely subwavelength-spaced elements with tunable impedance characteristics~\cite{Chen_2016_metasurfaces}. Reconfigurable Intelligent Surfaces (RISs) represent a key class of metasurface architectures integrated into the wireless environment without requiring active Radio-Frequency (RF) chains or internal feeds~\cite{RIS_overview}. By locally tailoring their electromagnetic response, RISs can dynamically manipulate impinging waves, dynamically reconfiguring the wireless channel to enhance coverage and sensing~\cite{RIS_SRE}. In contrast, waveguide-fed metasurface antennas, such as Dynamic Metasurface Antennas (DMAs)~\cite{Shlezinger2021Dynamic}, serve as active transceiver platforms. They realize densely sampled, near-continuous apertures through simple feed networks, eliminating complex power-splitting architectures and enabling highly energy efficient beam synthesis \cite{williams2023EM_DMA,davidsmith2017}. These antenna array architectures naturally scale to eXtremely Large Multiple-Input Multiple-Output (XL-MIMO) systems, with apertures spanning tens or even hundreds of wavelengths, thereby supporting near-field beamforming~\cite{zhang2022beam}, enhanced spatial multiplexing~\cite{Nlos_DMA}, and high-resolution sensing~\cite{gavras2025near}.

Realizing the latter functionalities with DMAs, in practice, demands physically-consistent models that explicitly account for strong mutual coupling; nonetheless, much of the metasurface-based literature remains coupling-unaware, relying on simplified models. Indicatively, in~\cite{zhang2022beam}, precise DMA-based spatial beam focusing for sum-rate maximization was designed, while online user tracking in multipath-dominant environments via a receiving DMA was presented in~\cite{Nlos_DMA} that outperforms fingerprinting baselines. Additionally, the authors in~\cite{gavras2025near} proposed codebook- and greedy-based DMA designs that attain accuracy comparable to fully digital arrays. In~\cite{NF_beam_tracking}, near-field beam tracking between a DMA-based base station and a mobile user was analyzed, characterizing the optimal achievable BeamForming (BF) gain and quantifying losses due to position mismatch. Only very recently, a coupling-aware design for a DMA-based bistatic sensing setup was presented in~\cite{gavras2025electromagnetics}, where its robustness to scatterer position uncertainty and clock asynchrony was assessed.

This paper focuses on $2$-Dimensional (2D) waveguide-fed metasurface antenna arrays~\cite{pulidomancera2018}, a highly energy efficient transceiver architecture where the entire aperture is excited via a single feed. Unlike conventional DMAs that employ per-column or per-row microstrips (or $1$-dimensional waveguides) which are driven by dedicated transceiver RF chains, 2D metasurface antennas eliminate the need for multiple feeding networks, while enabling dense and continuous aperture excitation. However, in such 2D waveguide-fed architectures, mutual coupling becomes more pronounced, as it extends beyond individual rows or columns. In this paper, we first present a model for the aperture of 2D metasurface antennas that is based on a coupled-dipole formalism, capturing both waveguide and free-space interactions, and derive a per-element passivity constraint for their magnetic polarizabilities. Then, we develop a electromagnetics-compliant framework for bistatic sensing incorporating 2D waveguide-fed metasurface arrays. In particular, capitalizing on the presented array response model, we derive the Position Error Bound (PEB) for multi-target sensing in the radiative near field, and formulate a PEB minimization problem that is rendered tractable via a Neumann-series approximation of the dipole response, yielding a convex design in the array's per-element resonance strengths. Monte Carlo simulations highlight the advantages of the presented coupling-aware design framework for XL-MIMO-enabled bistatic sensing, quantifying the impact of element placement and their total number in the PEB performance under strong coupling conditions.

% \textit{Notations:}
% Vectors and matrices are represented by boldface lowercase and uppercase letters, respectively. The transpose, Hermitian transpose, inverse, and positive semidefiteness of $\mathbf{A}$ are denoted as $\mathbf{A}^{\rm T}$, $\mathbf{A}^{\rm H}$, $\mathbf{A}^{-1}$, and $\mathbf{A}\succeq0$, respectively. $\mathbf{I}_{n}$, and $\mathbf{0}_{n}$ ($n\geq2$) indicate the $n\times n$ identity, and zeros' matrices, respectively. $[\mathbf{A}]_{i,j}$ is the $(i,j)$-th element of $\mathbf{A}$, whereas notation $i:j$ as a matrix row/column index indicates its respective $i$-th till the $j$-th elements. $\|\mathbf{A}\|$ represent $\mathbf{A}$'s Euclidean norm. $|a|$,  and $\Re\{a\}$ are respectively the amplitude and real part of complex scalar $a$, $\mathbb{C}$ is the complex number set, and $\jmath\triangleq\sqrt{-1}$ is the imaginary unit. $\mathbb{E}\{\cdot\}$ is the expectation operator and $\mathbf{x}\sim\mathcal{CN}(\mathbf{a},\mathbf{A})$ represents a complex Gaussian random vector with mean $\mathbf{a}$ and covariance matrix $\mathbf{A}$.

\vspace{-0.2 cm}
\section{Metasurface Response and System Modeling}\label{sec: resp_model}
We consider an XL antenna array implemented as a 2D waveguide-fed metasurface of height \(h\) filled with air, as depicted in~\cite[Fig.~2]{pulidomancera2018}, which is excited through a single feed, modeled as an electric line source. This architecture exhibits strong mutual coupling, since all its elements interact both through the guided mode and the free-space radiation. To account for these effects, we adopt~\cite{pulidomancera2018}'s coupled-dipole formalism, according to which the subwavelength size of each constituting metamaterial justifies its representation as a magnetic dipole~\cite{williams2023EM_DMA,pulidomancera2018,davidsmith2017}. In general, each element is characterized by a polarizability tensor that relates the induced $3$-dimensional magnetic field to the dipole moment vector. In this paper, we restrict our attention to elements with a single non-zero tensor component, consistent with \cite[Fig.~4]{pulidomancera2018}, where only the $x$-directed polarizability is significant. Without loss of generality, the dipoles are assumed to be aligned along the \(x\)-axis such that the only non-zero polarizability component is also $x$-directed, with the metasurface lying in the $xy$-plane at $z=0$. Accordingly, each $n$-th element ($n=1,2,\ldots,N$ with $N$ denoting the total number of metamaterials at the 2D waveguide-fed metasurface) is characterized by a scalar polarizability coefficient $\alpha_n$ rather than a full tensor.

The frequency dependence of the polarizability of each $n$-th element can be modeled by a Lorentzian-type resonance~\cite{davidsmith2017}:
\vspace{-0.1cm}
\begin{equation}
    \alpha_n(\omega) \triangleq \frac{F_n \omega^2}{\omega_{0,n}^2 - \omega^2 + j \Gamma_n \omega},\vspace{-0.1cm}
    \label{eq:polarizability}
\end{equation}
where $\omega$ is the operating angular frequency, $F_n$ is a real constant related to the resonance strength, and $\omega_{0,n}$ is the resonance frequency of the $n$-th element with $\Gamma_n$ being its damping factor. Although only the $x$-directed component of the polarizability is non-zero, the magnetic dipoles radiate fields with both $x$- and $y$-components, as shown in \cite[eq.~(A9)]{pulidomancera2018}. Nevertheless, only the $x$-component of the local magnetic field induces a dipole moment, which reduces the governing relations from vector to scalar form. Throughout this paper, all dipole moments and magnetic fields are therefore expressed in terms of their $x$-components.

Denoting by $m_n\in\Compl$ the magnetic dipole moment of each $n$-th element and by $H_{\text{loc},n}\in\Compl$ the local magnetic field at its position, their constitutive relation is given as follows:\vspace{-0.1cm}
\begin{equation}\label{eq: Local Field and Dipole Moment}
    m_n(\omega) \triangleq \alpha_n(\omega) H_{\text{loc},n}(\omega).\vspace{-0.1cm}
\end{equation}
In the remainder of the paper, we have dropped the \(\omega\) dependency to simplify the notation. The local field acting on the $n$-th element is given by the sum of the field from the source, $H_{0,n}$, and the scattered fields generated by all other dipoles:\vspace{-0.1cm}
\begin{equation}
    H_{\text{loc},n} \triangleq H_{0,n} + \sum_{j=1, j\neq n}^{N} [\mathbf{G}]_{n,j} m_j,\vspace{-0.1cm}
    \label{eq:Hloc_expanded_final}
\end{equation}
where $[\mathbf{G}]_{n,j}$ models the interaction between elements $j$ and $n$, while the self-term  interactions have been excluded. By defining $\mathbf{r}_n\triangleq[r_{x_n},r_{y_n},0]$ as the global position vector of each $n$-th element, the interaction matrix $\mathbf{G}\in\mathbb{C}^{N\times N}$ is:
\begin{equation}\label{eq: Total Green Function}
[\mathbf{G}]_{n,j} \triangleq 
\begin{cases}
G_{\text{WG}}(\mathbf{r}_n-\mathbf{r}_j)+G_{\text{FS}}(\mathbf{r}_n-\mathbf{r}_j), & n\neq j,\\
0, & n=j,
\end{cases}
\end{equation}
$\forall j=1,\ldots,N$, with $G_{\text{WG}}$ representing the coupling through the waveguide and $G_{\text{FS}}$ the coupling in the free space. It is noted that, for $x$-directed dipoles in a parallel-plate waveguide of height $h$, $G_{\text{WG}}$ is defined as follows:
\begin{equation}\label{eq: Waveguide Green}
\begin{split}
    G_{\text{WG}}(\mathbf{r}_n-\mathbf{r}_j)
    \triangleq & -\frac{\jmath k^2}{8h}\bigg[
        H^{(2)}_{0}\!\left(k\rho_{n,j}\right)\\
        & - \cos\bigl(2\psi_{n,j}\bigr)H^{(2)}_{2}\!\left(k\rho_{n,j}\right)
    \bigg],
\end{split}    
\end{equation}
where $H^{(2)}_{\nu}(\cdot)$ denotes the Hankel function of the second kind and order $\nu$\cite[eqs.~(V-14) and (V-15)]{balanis2016antenna}, $k$ is the propagation wavenumber in the waveguide, $\rho_{n,j}\triangleq\lvert\mathbf{r}_n-\mathbf{r}_j\rvert=\sqrt{(r_{x_n}-r_{x_j})^2+(r_{y_n}-r_{y_j})^2}$, and $\psi_{n,j}\triangleq\mathrm{atan}\left((r_{y_n}-r_{y_j})/(r_{x_n}-r_{x_j})\right)$. 
The dyadic Green's function that relates the magnetic moment to the magnetic field in free-space is given via \cite[eq.~(8.52)]{Novotny_Hecht_2006} and: \textit{i)} using the duality between electric and magnetic properties from \cite[Table~3.2]{balanis2012advanced}; \textit{ii)} accounting for the image of the dipole induced by the adjacent perfectly conducting surfaces; and \textit{iii)}  retaining only the $x$ to $x$ contribution of the dyadic Green, yields:
\begin{equation}\label{eq: Free Space Green}
\begin{split}
    G_{\text{FS}}(\mathbf{r}_n-\mathbf{r}_j) \triangleq & \left[
    \left(\frac{3}{k^2 \rho_{n,j}^2}+\frac{3\jmath}{k\rho_{n,j}}-1\right)\cos^2\left(\psi_{n,j}\right)\right.\\
    &\left.+\left(1-\frac{\jmath}{k\rho_{n,j}}-\frac{1}{k^2 \rho_{n,j}^2}\right)
    \right]\frac{k^2e^{-ik\rho_{n,j}}}{2\pi\rho_{n,j}}.
\end{split}
    \end{equation}
		
The feed is modeled as an electric line source with current \(I\) and its induced field \(H_{0,n}\) on the \(n\)-th element is given as~\cite{pulidomancera2018}:
\begin{equation}\label{eq: Excitation Field}
    H_{0,n}\triangleq\frac{\jmath k}{4}I \, H^{(2)}_{1}\!\left(k|\mathbf{r}_n-\mathbf{p}|\right)\sin\!\left(\mathrm{atan}\!\left( \frac{p_{y}- r_{y_n}}{p_{x}- r_{x_n}}\right)\right),
\end{equation}
where \(\mathbf{p}\triangleq[p_{x},p_{y},0]\) denotes the position of the source. To this end, we can compactly express the feed's contribution by introducing the coupling vector \(\mathbf{h}_f\in\mathbb{C}^{N \times 1}\) with \([\mathbf{h}_f]_{n}=\frac{\jmath k}{4}H^{(2)}_{1}\!\left(k|\mathbf{r}_n-\mathbf{p}|\right)\sin\!\left(\mathrm{atan}\!\left( \frac{p_{y}- r_{y_n}}{p_{x}- r_{x_n}}\right)\right)\) and the excitation-field vector $\mathbf{h}_0 \triangleq [H_{0,1},\ldots,H_{0,N}] \in \mathbb{C}^{N\times 1}$, yielding $\mathbf{h}_0 = \mathbf{h}_f I$. We also define the dipole moment vector as $\mathbf{m} \triangleq [m_1,\ldots,m_N] \in \mathbb{C}^{N\times 1}$ and the polarizability vector as $\boldsymbol{\alpha} \triangleq [\alpha_1,\ldots,\alpha_N] \in \mathbb{C}^{N\times 1}$. With these definitions, and since \eqref{eq:Hloc_expanded_final} holds for all metamaterial elements simultaneously, we can substitute \(H_{{\rm loc},n}\) therein with the respective dipole moments \(m_n\)'s via \eqref{eq: Local Field and Dipole Moment} and subsequently solve with respect to the dipole moment vector \(\mathbf{m}\), yielding the expression:
\begin{equation}\label{eq: Dipole Moment}
  \mathbf{m} = I\left(\operatorname{diag}(\boldsymbol{\alpha})^{-1} - \mathbf{G}\right)^{-1}\mathbf{h}_f .
\end{equation}
This expression provides a novel compact characterization of the coupled-dipole-based underlining  physics for 2D waveguide-fed metasurfaces with single-polarization elements.

\subsection{Power Conservation Constraint}
The constituent metamaterials do not perform amplification, implying that the magnetic polarizability of each dipole model must satisfy a passivity constraint (conservation of energy principle): the power supplied to each $n$-th dipole, $P_{\mathrm{sup}}$, cannot be smaller than its radiated power, $P_{\mathrm{rad}}$. While this principle has been considered for electric dipoles in free space \cite{Tretyakov2000,bohren2008absorption}, it is missing for magnetic dipoles in 2D waveguides despite  their usage in modeling metasurfaces~\cite{pulidomancera2018,davidsmith2017,williams2023EM_DMA}.  

Following the Poynting theorem~\cite[eq.~(1-76b)]{balanis2012advanced}, the supplied power can be derived by integrating over the effective volume \(V\) of the dipole\footnote{In its definition, the Poynting theorem is expressed with respect to the magnetic field vector, but since the magnetic current has only an \(x\)-component, the \(y\)-component of the magnetic field does not contribute power.}:
\begin{equation}\label{eq: Poynting Theorem}
P_{\mathrm{sup}} = -0.5\Re\left\{\int_{V} {M}_n(\mathbf{r})  {H}_{\mathrm{loc},n}^*(\mathbf{r})d\mathbf{r}\right\},
\end{equation}
where ${M}_n(\mathbf{r}) = j\omega\mu_0 {m}_n\delta(\mathbf{r}-\mathbf{r}_n)$ is the induced magnetic current density~\cite[eq.~(8.49)]{Novotny_Hecht_2006}, yielding: 
\begin{equation}\label{eq: Supplied Power}
P_{\mathrm{sup}} = -0.5\omega \mu_0\Im\{\alpha_n\}\left|{H}_{\mathrm{loc},n}\right|^2.
\end{equation}

On the other hand, to compute the radiated power, the Poynting theorem can be again used, but the induced magnetic field needs to be substituted by the scattered field of the \(n\)-th element, which is given as ${H}_{\mathrm{sc},n}(\mathbf{r})\triangleq {G}(\mathbf{r}-\mathbf{r}_n){m}_n$ with ${G}(\mathbf{r}-\mathbf{r}_n)={G}_{\mathrm{FS}}(\mathbf{r}-\mathbf{r}_n)+{G}_{\mathrm{WG}}(\mathbf{r}-\mathbf{r}_n)$. Finally, this yields:
\begin{equation}\label{eq: Radiated Power 1}
P_{\mathrm{rad}} = -0.5\omega\mu_0\left|\alpha_n\right|^2\left|\mathbf{H}_{\mathrm{loc},n}\right|^2\Im\left\{\mathbf{G}^{\mathrm{FS}}(0)+\mathbf{G}^{\mathrm{WG}}(0)\right\}.
\end{equation}
Evaluating the singularity term \(\Im\{\mathbf{G}^{\mathrm{FS}}(0)\}\) yields $-k^3/(3\pi)$~\cite{tretyakov2020magneticdipoles,williams2023EM_DMA}. This can be derived by taking the limit of~\eqref{eq: Free Space Green} at \(\mathbf{r}_n\to \mathbf{r}_j\) and using the Taylor expansion near zero of the term~\(\exp(-\jmath k\rho)\)~\cite[eq.~(4)]{tretyakov2020magneticdipoles}. For the latter term~$\Im\{\mathbf{G}^{\mathrm{WG}}(0)\}$, the singular terms of the Hankel functions are computed via~\cite[eqs.~(V-11) and (V-12)]{balanis2016antenna} yielding $-k^2/(8h)$. Hence, the radiated power is given as:  
\begin{equation}\label{eq: Radiated Power 2}
P_{\mathrm{rad}} = \tfrac{1}{2}\omega\mu_0|\alpha_n|^2|\mathbf{H}_{\mathrm{loc},n}|^2\!\left(\tfrac{k^3}{3\pi}+\tfrac{k^2}{8h}\right).
\end{equation}  
By imposing $|P_{\mathrm{rad}}|\leq|P_{\mathrm{sup}}|$, gives the passivity constraint:  
\begin{equation}\label{eq: constraint}
    \big|\Im\{\alpha_n^{-1}\}\big| \;\geq\; \frac{k^3}{3\pi}+\frac{k^2}{8h}.
\end{equation}

\subsection{System and Channel Models}
We consider a narrowband bistatic sensing system comprising a Transmitter (TX) equipped with a 2D waveguide-fed metasurface antenna array and a conventional multi-antenna Receiver (RX), which wishes to localize \(U\) nearby targets, modeled as Scattering Points (SPs). The TX array comprises \(N\) metamaterials driven by a single excitation feed, whereas the RX, centered at \(\p_{\rm RX}\in\mathbb{R}^3\), includes \(M\) RF chains connected one-to-one to \(M\) antennas with half-wavelength spacing (i.e., \(\lambda/2\)), enabling digital combining of the received signals encompassing spatial information about the \(U\) targets. This combining operation is modeled by the matrix $\mathbf{W}_{\rm RX} \in \mathbb{C}^{M \times M}$, and we assume that $U\leq{\rm \min}\{N, M\}$ to ensure the feasibility of the BF design for target sensing.

Assuming operation in the radiative near-field~\cite{NF_beam_tracking}, the $M$-element baseband received signal at the RX side is expressed as $\mathbf{y} \triangleq \W_{\rm RX}^{\rm H}\H_{\rm R}\m + \mathbf{n}$, where $\mathbf{n}\sim\mathcal{CN}(\mathbf{0},\sigma^2\mathbf{I}_{M})$ denotes the Additive White Gaussian Noise (AWGN). By substituting the TX signal $\m$ from~\eqref{eq: Dipole Moment}, the received signal becomes $\mathbf{y} = I\W_{\rm RX}^{\rm H}\H_{\rm R}\big(\operatorname{diag}(\boldsymbol{\alpha})^{-1} - \mathbf{G}\big)^{-1}\mathbf{h}_f + \mathbf{n}$. Concatenating the received signals from \(T\) pilot transmissions and denoting the transmitted pilot sequence as $\mathbf{i} \triangleq [I_1,\ldots,I_T]^{\rm T} \in \Compl^{T\times 1}$, the overall $M\times T$ matrix with the received pilots is given by:
\begin{equation}\label{eq: Received Signal_matrix}
    \mathbf{Y} \triangleq \W_{\rm RX}^{\rm H}\H_{\rm R}\big(\operatorname{diag}(\boldsymbol{\alpha})^{-1} - \mathbf{G}\big)^{-1}\mathbf{h}_f\mathbf{i}^{\rm T} + \mathbf{N},
\end{equation}
where $\mathbf{N} \triangleq [\mathbf{n}(1),\ldots,\mathbf{n}(T)] \in \Compl^{M\times T}$ with each $\mathbf{n}(t)\sim\mathcal{CN}(\mathbf{0},\sigma^2\mathbf{I}_{M})$ denoting the AWGN over all pilot transmissions. Furthermore, for each transmitted pilot \([\i]_t\), it holds that $\|[\i]_t\|^2=P_{\rm TX}$ with \(P_{\max}\) being the available transmit power budget, and $\H_{\rm R}\in\Compl^{M\times N}$ represents the composite round-trip channel matrix, which is modeled as follows:\vspace{-0.1cm}
%Furthermore, $\s\in\Compl^{1\times T}$ are the DL pilots vector, with $\|\m[\s]_t\|^2\leq P,\,\forall \, t=1,2,\ldots,T$, and $\H_{\rm R}\in\Compl^{M\times N}$ is the end-to-end composite channel matrix, which is modeled as follows:\vspace{-0.5cm}
\begin{align}\label{eq:H_R}
    \H_{\rm R} \triangleq \sum_{u=1}^U \beta_u \a_{\rm RX}\left(\p_u\right)\a_{\rm TX}^{\rm H}\left(\p_u\right),
\end{align}
where $\beta_u$ is the reflection coefficient of each $u$-th SP, which is modeled as a
point scatterer positioned between the TX and RX at the point $\p_u\triangleq[x_u,y_u,z_u]$.
% Due to near-field propagation assumption, each SP is modeled as a
% point source \cite{} positioned between the TX and RX at the points $\p_u\triangleq[x_u,y_u,z_u]$, $\forall u=1,\ldots,U$. 
The focusing vectors at the TX and the RX are denoted by $\a_{\rm TX}(\cdot) \in \Compl^{N\times 1}$ and $\a_{\rm RX}(\cdot) \in \Compl^{M \times 1}$, respectively. Note that both nodes in the considered bistatic sensing setup are assumed to know each other's 3D positions and to be synchronized. Under these assumptions, the RX can reliably suppress the direct line-of-sight (LoS) component from the TX (e.g., via direct-path cancellation)~\cite{saini2003direct}. We henceforth use the definition of the auxiliary variable \(\bar{\m} \triangleq \big(\operatorname{diag}(\boldsymbol{\alpha})^{-1} - \mathbf{G}\big)^{-1}\mathbf{h}_f\) to distinguish the contribution of the source current $I$ from the dipole moments, i.e., \(\m = \bar{\m}I\).

By solving the radiation problem in the region outside the waveguide, the electric field \(E_{{\rm sc},n}\) radiated by each \(n\)-th metamaterial element can be expressed as a function of its magnetic moment \(m_n\). While the authors in \cite{pulidomancera2018} derive this field in spherical coordinates under a far-field approximation, we instead adopt a radiative near-field model. To this end, we eliminate only the terms divided by the squared distance \(R^2_n\) between the \(n\)-th element and the observation point. The resulting expression is then recasted in Cartesian coordinates and we analyze only the \(z\)-component of \(E_{{\rm sc},n}\), which is the dominant term in this case. 
%The electric field is acquired from the magnetic field components as \(\nabla_{\mathbf{r}}\times {H}_{{\rm sc},n}(\mathbf{r})=-\jmath \omega \epsilon_0 E_{{\rm sc},n}(\mathbf{r})\).
Following~\cite[eqs.~(12) and (13)]{pulidomancera2018}, transforming the electric field from spherical to Cartesian via~\cite[eq.~(II-13a)]{balanis2012advanced} and keeping only the \(z\)-component, the electric field at the SPs' positions $\p_u$ $\forall u$ is given as:
\begin{equation}\label{eq: E-field of n-th metamaterial}
    E_{{\rm sc},n}(\p_u)=-k^2 \eta \sin\left(\theta_{{\rm e},n,u}\right)\sin\left(\theta_{{\rm a},n,u}\right)\frac{e^{-\jmath k R_{n,u}}}{2 \pi R_u} m_n,
\end{equation}
where \(\eta\) is the free-space impedance, \(R_{n,u}\) is the radial distance between $\p_u$ and \(\mathbf{r}_n\), and \(\theta_{{\rm e},n,u}\triangleq{\rm acos}\left(({z_u - r_{{z}_n}})/{R_n}\right)\) as well as \(\theta_{{\rm a},n,u} \triangleq {\rm atan}\left( ({y_u- r_{y_n}})/({x_u- r_{x_n}})\right)\) are respectively the elevation and azimuth Angles of Departure (AoDs) between them. Therefore, since $\a_{\rm TX}(\cdot)$ is the focusing vector modeling the transfer of the electric field between the TX array and the observation point $\p_u$ $\forall u$, it can be written as follows: 
\begin{align}\label{eq: Focusing Vector}
    \a_{\rm TX}(\p_u) = \frac{-k^2\eta}{2 \pi R_u}&\left[\color[rgb]{0,0,0}\sin(\theta_{{\rm e},1,u})\sin(\theta_{{\rm a},1,u})e^{\jmath k R_{1,u}}\right.,\\&
    \hspace{0.1cm}\left.\ldots,\sin(\theta_{{\rm e},N,u})\sin(\theta_{{\rm a},N,u})e^{\jmath k R_{N,u}}\right],\nonumber 
\end{align}
where \(R_{u}\) denotes the distance between the center of the TX and \(\p_u\)%\footnote{Here we assume that the radial distance from the array center to the observation point is taken to be approximately equal to the distance from each element; i.e., $R_u\approx R_{1,u}\approx\ldots\approx R_{N,u}$, $\forall u$.}
, while \(R_{n,u},\theta_{{\rm e},n,u}\), and \(\theta_{{\rm a},n,u}\) $\forall n,u$ follow from the previous definitions. The RX focusing vector \(\a_{\rm RX}(\cdot)\) is defined analogously to \cite{gavras2025electromagnetics}, using the radial distances \(B_{m,u}\) between the \(m\)-th RX antenna and the \(u\)-th SP, together with the corresponding elevation and azimuth Angles of Arrival (AoAs) \(\psi_{{\rm e},m,u}\) and \(\psi_{{\rm a},m,u}\) \(\forall m=1,\ldots,M\) and \(\forall u=1,\ldots,U\).

\section{Proposed Design for Bistatic Sensing}
In this section, we first derive the PEB metric quantifying the multi-target sensing capability of the considered bistatic sensing system setup incorporating a 2D waveguide-fed metasurface TX. Next, capitalizing on the physically consistent metasurface response model in Section~\ref{sec: resp_model}, we present an optimization framework of its per-element resonance strengths. 

\subsection{Cram\'{e}r-Rao Bound (CRB) Derivation}
Let us introduce $\boldsymbol{\xi}\triangleq[\boldsymbol{\theta}_{\rm a}^{\rm T},\boldsymbol{\theta}_{\rm e}^{\rm T},\boldsymbol{\psi}_{\rm a}^{\rm T},\boldsymbol{\psi}_{\rm e}^{\rm T},\boldsymbol{\rho}^{\rm T},\boldsymbol{\upsilon}^{\rm T},\boldsymbol{\beta}^{\rm T}]^{\rm T}\in\mathbb{R}^{L\times 1}
$, where $L=U(3N+3M+2)$ and $\widetilde{\boldsymbol{\xi}}\triangleq[\p^{\rm T},\boldsymbol{\beta}^{\rm T}]^{\rm T}\in\mathbb{R}^{5U\times 1}$ include respectively the unknown channel and location parameters, with $\boldsymbol{\theta}_{\rm a}\triangleq[\theta_{{\rm a},1,1},\ldots,\theta_{{\rm a},N,U}]$, $\boldsymbol{\theta}_{\rm e}\triangleq[\theta_{{\rm e},1,1},\ldots,\theta_{{\rm e},N,U}]$, $\boldsymbol{\psi}_{\rm a}\triangleq[\psi_{{\rm a},1,1},\ldots,\psi_{{\rm a},M,U}]$, $\boldsymbol{\psi}_{\rm e}\triangleq[\psi_{{\rm e},1,1},\ldots,\psi_{{\rm e},M,U}]$, $\boldsymbol{\rho}\triangleq[R_{1,1},\ldots,R_{N,U}]$, $\boldsymbol{\upsilon}\triangleq[B_{1,1},\ldots,B_{M,U}]$, $\boldsymbol{\beta}\triangleq[\Re\{\beta_1\},\ldots,\Re\{\beta_U\},\Im\{\beta_1\},\ldots,\Im\{\beta_U\}]$, and $\p\triangleq[\p_1,\ldots,\p_U]^{\rm T}\in\mathbb{R}^{3U\times 1}$. It is evident from~\eqref{eq: Received Signal_matrix}'s inspection that the received signal at the outputs of the RX's RF chains modeled via \(\Y\) has \(E[\Y]= \W_{\rm RX}^{\rm H}\H_{\rm R}\bar{\m}\i^{\rm T}\) and \(E[(\Y-E[\Y])(\Y-E[\Y])^{\rm H}]=\sigma^2\I_{M} \). Then, each $(i,j)$-th element (with $i,j=1,\ldots,L$) of the Fisher Information Matrix (FIM) $\J \in \Compl^{L\times L}$ associated with the channel parameters' vector $\boldsymbol{\xi}$ can be computed as follows \cite{kay1993fundamentals}:%\vspace{-0.1cm}
\begin{align}\label{eq: FIM}
    [\J]_{i,j} \triangleq \frac{2T}{\sigma^2}{\rm Tr}\left\{\Re\left\{\!\frac{\partial {\boldsymbol{\mu}}^{\rm H}}{\partial[\boldsymbol{\xi}]_i}\frac{\partial {\boldsymbol{\mu}}}{\partial[\boldsymbol{\xi}]_j}\!\right\}\right\}.
\end{align}
where ${\boldsymbol{\mu}}\triangleq\W_{\rm RX}^{\rm H}\H_{\rm R}\bar{\m}$. Focusing on the SPs' position estimation, we next deploy the transformation matrix $\T\in\mathbb{R}^{L\times5U}$, which can be expressed as a Jacobian with $[\T]_{i,j}=\partial[{\boldsymbol{\xi}}]_i/\partial[\widetilde{\boldsymbol{\xi}}]_j$ $\forall i,j$, to derive the FIM $\widetilde{\J}\in\Compl^{5U\times5U}$ of the locations' parameter vector $\widetilde{\boldsymbol{\xi}}$ as $\widetilde{\J} = \T^{\rm T}\J\T$. Recall that the RX's position is not included in $\widetilde{\boldsymbol{\xi}}$, since it is assumed to be known at the TX, and vice versa. 

By using the submatrix definitions $\widetilde{\J}_{\p\p} \triangleq [\widetilde{\J}]_{1:3U,1:3U}$, $\widetilde{\J}_{\p\boldsymbol{\beta}} \triangleq [\widetilde{\J}]_{1:3U,3U+1:5U}$, and $\widetilde{\J}_{\boldsymbol{\beta}\boldsymbol{\beta}}\triangleq[\widetilde{\J}]_{3U+1:5U,3U+1:5U}$, and the properties of the Schur's complement, we can assess the accuracy of the SP positions' estimates through the PEB sensing performance metric, which is computed as follows:
\begin{align}\label{eq: PEB_schur}
    {\rm PEB}\left(\bar{\m};\widetilde{\boldsymbol{\xi}}\right) = \sqrt{{\rm Tr}\left\{\left[\widetilde{\J}_{\p\p}-\widetilde{\J}_{\p\boldsymbol{\beta}}\widetilde{\J}_{\boldsymbol{\beta}\boldsymbol{\beta}}^{-1}\widetilde{\J}_{\p\boldsymbol{\beta}}^{\rm T}\right]^{-1}\right\}}.
\end{align}
\vspace{-0.62cm}

\subsection{Problem Formulation}
Our objective is to optimize the resonance strengths $F_n$ $\forall n$ of the considered 2D waveguide-fed metasurfaces in order to maximize the bistatic sensing accuracy. We assume a common resonance frequency \(\omega_0\) and operation in resonance, i.e., $\omega_{0,n}=\omega_0=\omega$ $\forall n$, thus, \([\boldsymbol{\alpha}]_n= -\jmath F_n\omega_0/\Gamma_n\). However, direct optimization under the proposed physically consistent model in~\eqref{eq: Dipole Moment} is challenging due to the inverse of the sum involving the inverse polarizability and the mutual-coupling matrix. To this end, we have chosen to first relax the initial problem to the case where the dipole moment vector \(\bar{\m}\) is an optimizable variable with only a constraint on its squared norm. To remain close to the actual feasible space, the constraint is set so as to meet the following reasoning. Assume weak mutual coupling, then, \(\bar{\m}\approx {\rm diag}(\boldsymbol{\alpha})\h_f\). Furthermore, since the polarizabilities are purely imaginary, using \eqref{eq: constraint} and the fact that the entries of \(\h_f\) have amplitude less than \(1\), yields \(|[\bar{\m}]_n|\leq C\) with \(C\triangleq\frac{k^3}{3\pi}+\frac{k^2}{8h}\). We further relax this constraint to a squared-norm one to enable us to handle the non-convexity of the per-element amplitude constraints, and particularly write \(\|\bar{\m}\|^2_{\rm F}\leq N C^2\). The latter brings us to the metasurface design optimization objective for bistatic sensing:

\begin{align}
        \mathcal{P}_1:\,\nonumber&\underset{\bar{\m}}{\min} \,\,{\rm PEB}\left(\bar{\m};\widetilde{\boldsymbol{\xi}}\right)\,\text{\text{s}.\text{t}.}\,\, \|\bar{\m}\|^2\leq N C^2.
\end{align}
However, even in this formulation, the PEB is non-convex w.r.t. \(\bar{\m}\). It can be though deduced from~\eqref{eq: FIM} that the FIM matrix $\J$, and thus the PEB, can be reformulated to be linear in terms of the matrix $\M\triangleq\!\bar{\m}\bar{\m}^{\rm H}$.
%
% Direct optimization under the physically consistent model in \eqref{eq: Dipole Moment} is challenging due to the inverse of the sum involving the polarizability and mutual-coupling matrices. Following the procedure in Sec.~II of \cite{gavras2025electromagnetics}, we approximate the dipole moment via a Neumann series expansion up to second order as $\bar{\m} \approx -\G^{-1}\mathrm{diag}(\boldsymbol{\alpha})^{-1}\G\h_f$, which substantially simplifies \eqref{eq: Dipole Moment} for optimization. The optimization objective can be mathematically expressed as follows:
% \begin{align}
%         \mathcal{P}_1:\,\nonumber&\underset{\substack{\{F_n\}_{n=1}^N}}{\min} \,\,{\rm PEB}\left(\bar{\m};\widetilde{\boldsymbol{\xi}}\right)\,\text{\text{s}.\text{t}.}\,\, \|\bar{\m}\|^2\leq P_{\rm max}.
% \end{align}
% In this formulation, even with the approximated dipole moment model, direct optimization of $F_n$, $\forall n$ is cumbersome. It can be deduced from $\eqref{eq: FIM}$, that the FIM matrix $\J$ and thus the PEB can be reformulated to be linear in terms of the matrix $\M\triangleq\!\bar{\m}\!\bar{\m}^{\rm H}$. 
%
Following a similar proof to~\cite{keskin2022optimal}, it can be shown that the optimal covariance of $\M$ in $\mathcal{P}_1$ follows the structure $\M_{\rm opt}=\U_{\rm TX}\boldsymbol{\Lambda}\U_{\rm TX}^{\rm H}$, with $\boldsymbol{\Lambda}\in\Compl^{4U\times 4U}$ being a positive semidefinite diagonal matrix and $\U_{\rm TX}\triangleq[\A_{\rm TX},\A_{{\rm TX},x},\A_{{\rm TX},y},\A_{{\rm TX},z}]\in\Compl^{N\times 4U}$, where $\A_{\rm TX}\triangleq[\a_{\rm TX}(\p_1),\ldots,\a_{\rm TX}(\p_U)]\in\Compl^{N\times 1}$ and $\A_{{\rm TX},x},\A_{{\rm TX},y}, \A_{{\rm TX},z}\in\Compl^{N\times 1}$ correspond to the respective derivative subspaces according to each spatial dimension of $\p_u$ $\forall u$. Following these steps and utilizing the Schur's complement as in~\eqref{eq: PEB_schur}, $\mathcal{P}_1$ can be simplified as follows:
\begin{align}
        \nonumber&\underset{\substack{\Z,\boldsymbol{\Lambda}}}{\min} \,\,{\rm Tr}\{\Z^{-1}\}\\
        &\nonumber\text{\text{s}.\text{t}.}\,\nonumber
    \begin{bmatrix}
    \widetilde{\J}_{\p\p}(\M)-\Z & \widetilde{\J}_{\p\boldsymbol{\beta}}(\M) \\
    \widetilde{\J}_{\p\boldsymbol{\beta}}^{\rm T}(\M) & \widetilde{\J}_{\boldsymbol{\beta}\boldsymbol{\beta}}(\M)
    \end{bmatrix}\succeq0,\,\Z\succeq0,\, \boldsymbol{\Lambda}\succeq 0,
    \\&\nonumber\qquad {\rm Tr}\{\M\}\leq NC^2,\,\M=\U_{\rm TX}\boldsymbol{\Lambda}\U_{\rm TX}^{\rm H},
\end{align}
where $\Z\in\Compl^{3U\times3U}$ is an auxiliary optimization variable and the FIM submatrices $\widetilde{\J}_{\p\p},\widetilde{\J}_{\p\boldsymbol{\beta}}$, and $\widetilde{\J}_{\boldsymbol{\beta}\boldsymbol{\beta}}$ are rewritten in terms of $\M$ stemming from \eqref{eq: FIM}. This formulation results in a convex semidefinite program which can be efficiently solved using standard off-the-shelf convex optimization solvers. An approximate solution $\bar{\m}_{\rm opt}\in\Compl^{N\times 1}$ can be acquired by considering \(\M_{\rm opt}\) to be the covariance matrix of the solution and sampling from it, or via a rank-one approximation \cite{gavras2025electromagnetics}. 

We now proceed to retract $\m_{\rm opt}$ to the feasible set by solving a least-squares problem between \(\bar{\m}_{\rm opt}\) and \(\bar{\m}=\big(\operatorname{diag}(\boldsymbol{\alpha})^{-1} - \mathbf{G}\big)^{-1}\mathbf{h}_f\). To ensure convexity for this problem, we use a Neumann-series approximation for the inverse resulting in\footnote{We first apply the Woodbury formula yielding ${\rm diag}(\boldsymbol{\alpha})^{-1}-{\rm diag}(\boldsymbol{\alpha})^{-1}(-{\rm diag}(\boldsymbol{\alpha})\G^{-1}+\I_{\rm N})^{-1}\h_f$, then, use $(-{\rm diag}(\boldsymbol{\alpha})\G^{-1}+\I_{\rm N})^{-1}\approx\I_N+\sum_{p=1}^{\infty}(-1)^p(-{\rm diag}(\boldsymbol{\alpha})\G^{-1})^p$ and truncate up to $p=2$.} \(\bar{\m} \approx -(\G^{-1}+\G^{-1}\mathrm{diag}(\boldsymbol{\alpha})^{-1}\G^{-1})\h_f\) \cite{gavras2025electromagnetics}. Capitalizing on this approximation, we focus on solving:
\begin{align}
    &\nonumber\mathcal{P}_2:\underset{\substack{\{F_n\}_{n=1}^{N}}}{\min} \,\,\left\|\bar{\m}_{\rm opt}+(\G^{-1}+\G^{-1}\mathrm{diag}(\boldsymbol{\alpha})^{-1}\G^{-1})\h_f\right\|^2\,
    \\&\nonumber\qquad\,\text{\text{s}.\text{t}.}\, 0< F_n\leq\frac{\Gamma_n}{C\omega}\,\forall n, 
\end{align}
where the per-element resonance strength constraint follows directly from \eqref{eq: constraint}. This formulation yields a convex quadratic problem since the terms $1/F_n$ $\forall n$ are linearly dependent; consequently, $\mathcal{P}_2$ can be solved efficiently using standard optimization solvers.

\begin{figure}[!t]
	\begin{center}
	\includegraphics[scale=0.6]{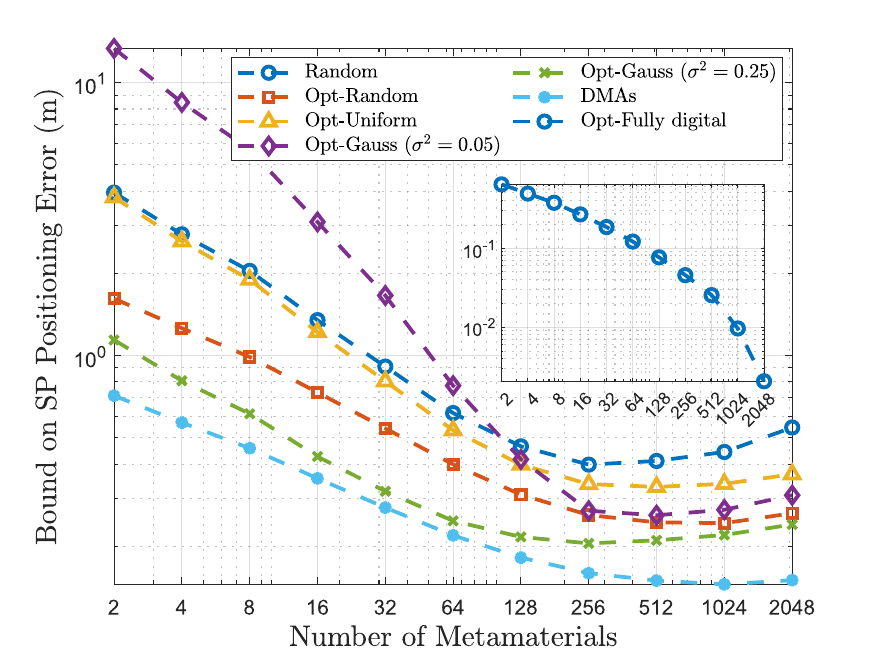}
	\caption{\small{PEB performance with the proposed bistatic sensing design considering $U=2$ SPs versus the number of metamaterials under different placements of them across the 2D metasurface aperture.}} 
    \vspace{-0.4cm}
	\label{fig: PEB}
	\end{center}
\end{figure}

\section{Numerical Results and Discussion}\label{Sec: Numerical}
In this section, we evaluate the performance of the proposed design of 2D waveguide-fed metasurface antenna arrays for bistatic sensing via Monte Carlo simulations. In particular, the target sensing accuracy has been assessed as a function of: \textit{i}) the number of metamaterial elements $N$; as well as \textit{ii}) their spatial placement on the aperture. We have assumed operation at \(20\) GHz with a coherent channel block comprising \(T=100\) pilot transmissions, and \(U=2\) SPs were located at the unknown positions \(\p_1=[5.4,\,5.3,\,4]\) and \(\p_2=[7.1,\,3.5,\,5.25]\). The metasurface was a \(0.5{\rm\,m}\times0.5{\rm\,m}\) square panel, the feed was placed at the center of the aperture, and the waveguide height was set to \(h=\lambda/5\). Its lements were assumed to satisfy a minimum spacing of \(\lambda/4\). The RX was assumed to employ \(M=16\) antennas centered at \(\p_{\rm RX}=[10,\,5,\,5]\), with \(\W_{\rm RX}\) drawn at random from a discrete Fourier transform codebook. The obtained results were averaged over \(500\) Monte Carlo trials. The AWGN variance was set to \(\sigma^2=-80\) dBm, the coefficients \(\beta_u\) in \eqref{eq:H_R} were drawn with unit magnitude \(\forall u\in\{1,2\}\), and the TX power was set as $P_{\rm TX}=1$ dBm.

Figure~\ref{fig: PEB} reports the bistatic PEB performance versus $N$, ranging from sparse apertures to thousands of elements (i.e., the XL MIMO regime). For a fair comparison, we have evaluated a variety of element-placement strategies and array architectures: \textit{i}) random placement within the aperture, with and without resonance strength optimization; \textit{ii}) a uniform grid of equally spaced elements; \textit{iii}) Gaussian-clustered placement centered on the aperture with variances \(\sigma^2=0.05\) and \(\sigma^2=0.25\); \textit{iv}) a physics-consistent DMA baseline following~\cite{gavras2025electromagnetics} under the same aperture and element count but comprising \(4\) RF chains; and \textit{v}) the proposed fully digital benchmark \(\m_{\rm opt}\) with \(N\) RF chains. As observed, as $N$ grows, the PEB initially decreases due to higher resolution BF. However, for \( 256\leq N\leq512\), PEB degrades due to: \textit{i}) intensified mutual coupling from reduced inter-element spacing; and \textit{ii}) weaker radiation of peripheral elements since the excitation field attenuates away from the centrally located source. Among placements, high-variance Gaussian layouts yield the lowest PEB, likely thanks to reduced mutual coupling compared to the low-variance Gaussian, and the more dense placement near the center compared to random and uniform. The DMA and fully digital baselines outperform the proposed 2D waveguide-fed design, since the DMA induces less mutual coupling via multiple 1D waveguides with dedicated RF chains, while the fully digital array is uncoupled and enables ideal precoding. Yet, the proposed architecture only requires a single RF chain, thus, drastically reducing power consumption and retaining comparable accuracy to DMA.

%As expected, increasing the number of metamaterial elements lowers the PEB by enriching the effective spatial sampling; however, beyond roughly 256–512 elements the trend reverses because the spacing between the elements shrinks, and thus mutual coupling intensifies. Gaussian-distributed layouts with larger variance yield the best accuracy, likely due to broader edge coverage and reduced coupling correlation. The DMA and fully digital baselines outperform the proposed 2D waveguide-fed design due to the DMA arranging elements along multiple 1D waveguides with dedicated RF chains, which limits cross-waveguide coupling, while the fully digital array assumes independent elements with no coupling, enabling ideal precoding. Note that the proposed 2D waveguide architecture achieves comparable performance with the benchmarks while it is fed from a single excitation feed.

\section{Conclusions}
This paper presented a physics-consistent framework for bistatic sensing using a 2D waveguide-fed metasurface antenna array. A coupled-dipole model capturing both waveguide and free-space interactions was established together with a passivity constraint for the array's magnetic polarizabilities, which were incorporated into an CRB-based sensing optimization formulation having the metamaterials' resonance strengths as the free parameters. Our simulations for near-field conditions showcased the importance of appropriate element placement and mutual-coupling effects in the PEB performance.

\bibliographystyle{IEEEtran}
\bibliography{ms}
\end{document}

%% file: Definitions.tex
\renewcommand{\a}{\mathbf{a}}

\newcommand{\h}{\mathbf{h}}
\renewcommand{\i}{\mathbf{i}}

\newcommand{\m}{\mathbf{m}}

\newcommand{\p}{\mathbf{p}}

\newcommand{\A}{\mathbf{A}}

\newcommand{\G}{\mathbf{G}}
\renewcommand{\H}{\mathbf{H}}
\newcommand{\I}{\mathbf{I}}
\newcommand{\J}{\mathbf{J}}

\newcommand{\M}{\mathbf{M}}

\newcommand{\T}{\mathbf{T}}
\newcommand{\U}{\mathbf{U}}

\newcommand{\W}{\mathbf{W}}

\newcommand{\Y}{\mathbf{Y}}
\newcommand{\Z}{\mathbf{Z}}

\newcommand{\Compl}{\mbox{$\mathbb{C}$}}

\renewcommand{\Im}{\mathrm{Im}}

\renewcommand{\Re}{\mathrm{Re}}